\documentclass{aa}
\usepackage{txfonts}
\usepackage{graphicx}
\usepackage{natbib}
\bibpunct{ (}{)}{;}{a}{}{,}

\def\etal{et  al.\ }
\def\araa{{Ann.\ Rev.\ Astron.\ Ap.}}

\def\apj{ApJ}
\def\apjl{{ApJ\ (Lett.)}}
\def\apjs{{ApJ\ Suppl.}}

\def\aa{{A\&A}}
\def\aap{{A\&A}}
\def\mnras{{MNRAS}}

\def\keV {\rm keV}
\def\msol{{\rm M_{\odot}}}

\def \rv {R_{200}}
\def \mv {M_{200}}

\def\nh{{N_{\rm H}}}
\def\kT{{\rm k}T}

\def\rhoc{\rho_{\rm c}}

\def\Ts{T_{\rm s}}

\def\Tem{T_{\rm em}}
\def\Tm{T_{\rm m}}
\def\Tsl{T_{\rm sl}}

\def\MT{$M$--$T$}
\def\MdT{$M_{\delta}$--$T$}

\def\betamodel{$\beta$-model}
\def \etal {et al.\ }
\def \pks {\hbox{PKS 0745-191}}

\def \xmm {\hbox{\it XMM-Newton}}
\def \asca {\hbox{\it ASCA}}
\def \rosat {\hbox{\it ROSAT}}
\def \sax {\hbox{\it BeppoSAX}}
\def \chandra {\hbox{\it Chandra}}

\def\lesssim{\mathrel{\hbox{\rlap{\hbox{\lower4pt\hbox{$\sim$}}}\hbox{$<
$}}}}
\def\gtrsim{\mathrel{\hbox{\rlap{\hbox{\lower4pt\hbox{$\sim$}}}\hbox{$>$
}}}}

\begin{document}
\title{The structural and scaling properties of nearby galaxy clusters
- \\
    II. The \MT\ relation}

              \author{M. Arnaud \inst{1},
              E. Pointecouteau \inst{1} and
                      G.W. Pratt \inst{2}}
\offprints{M. Arnaud, \email{marnaud@discovery.saclay.cea.fr}}

           \institute{$^1$ CEA/DSM/DAPNIA Service d'Astrophysique, C.E.
Saclay, L'Orme des Merisiers,
Bat. 709, F-91191 Gif sur Yvette, France\\
           $^2$ MPE, Giessenbachstra{\ss}e, 85748 Garching, Germany}

\date{Received: February 10, 2005; Accepted: June 19, 2005}

\abstract{Using a sample of ten nearby ($z\lesssim 0.15$), relaxed
galaxy clusters in the temperature range $[2-9]~\keV$, we have
investigated the scaling relation between the mass at various density
contrasts ($\delta=2500,1000,500,200$) and the cluster
temperature. 
The masses are derived from NFW-type model fits to
  mass profiles, obtained under the hydrostatic assumption
using precise measurements, with \xmm, at least down to $\delta=1000$.
The logarithmic slope of the $M$--$T$ relation is well
constrained and is the same at all $\delta$, reflecting the
self-similarity of the mass profiles. At $\delta=500$, the slope of
the relation for the sub-sample of hot clusters ($\kT>3.5~\keV$) is
consistent with the standard self-similar expectation: $\alpha=
1.49\pm0.15$. The relation steepens when the whole sample is
considered: $\alpha=1.71\pm0.09$. The normalisation of the relation
is discrepant (by $\sim30$ per cent), at all density contrasts, with the
prediction from purely gravitation based models.  Models that take
into account radiative cooling and galaxy feedback are generally in
better agreement with our data.  We argue that remaining
discrepancies, in particular at low $\delta$, are more likely due to
problems with models of the ICM thermal structure rather than to
an incorrect estimate of the mass from X-ray data.

\keywords{Cosmology: observations,  Cosmology: dark
      matter, Galaxies: cluster: general, (Galaxies) Intergalactic  
medium, X-rays: galaxies: clusters}}

\authorrunning{Arnaud \etal }
\titlerunning{The $M-T$ relation}
\maketitle

%
%________________________________________________________________

\section{Introduction}
\label{sec:intro}

 From a theoretical
point of view, galaxy clusters are characterised by their mass.  Models  
of
structure formation predict the space density, spatial distribution
and physical properties of clusters (internal structure, radius,
temperature, luminosity, etc) as a function of mass and redshift
\citep[see][for a review]{bertschinger98}.  However, observationally,
the mass is not easily measured, and the observed scaling relations
are in fact expressed in terms of the temperature $T$, rather than the
mass $M$.  These scaling relations are important sources of
information on the physics of cluster formation
\citep[e.g.][]{voit03}. For the information to be complete, we must  
determine the
\MT\ relation itself, which provides the missing link between the gas
properties and the mass. Furthermore, measures of the cosmological
parameters, such as $\sigma_8$, $\Omega_{\rm m}$ and $w$, from cluster
abundance or spatial distribution, rely heavily on this relation to
link the mass to the X--ray observables available from X-ray cluster
surveys. The present error on the value of $\sigma_8$, as determined
from X--ray observations, is dominated by uncertainty on the \MT\
relation
\citep{pierpaoli03,viana03,henry04}, and a precise calibration of this
relation is mandatory if we want to do `precision' cosmology with clusters
\citep{borgani03}.

The average temperature is expected to be closely related to the mass,
via the virial theorem. We can first define $M_{\delta}$ as the mass
within the radius $R_{\delta}$, inside which the mean mass density is
$\delta$ times the critical density, $\rho_{\rm c}(z)= 3 h(z)^2{\rm
H_0} /8\pi{\rm G}$, at the cluster redshift. We then expect $h(z)
M_{\delta} = A(\delta) T^{3/2}$, {\it if} clusters are in hydrostatic
equilibrium {\it and} they obey self-similarity. Here $h(z)$ is the  
Hubble
constant normalised to its local value and $A(\delta)$ depends on the
internal structure.  The above relation is remarkably well verified by
adiabatic numerical simulations, down to $\delta\sim 200$, which
roughly corresponds to the virialised part of clusters
\citep[e.g.][]{evrard02}.

For relaxed clusters, the mass can be derived from X-ray observations
of the gas density and temperature profile and the hydrostatic
equilibrium equation. In recent years, a sustained observational
effort to measure the local \MT\ relation has been undertaken using
\rosat, \asca\ and \sax, but no definitive picture has yet emerged. It
is unclear whether the mass scales as $T^{3/2}$ as expected
\citep{horner99,ettori02,castillo03}; or if this is true only in the  
high mass
regime ($\kT\gtrsim4~\keV$), with a steepening at lower mass
\citep{nevalainen00,xu01,finoguenov01}; or even if the slope is higher
over the entire mass range \citep{sanderson03}. The derived
normalisations of the \MT\ relation derived from \asca\ data are
generally lower than predicted by adiabatic numerical simulations
\citep[e.g.][]{nevalainen00,finoguenov01}, typically $40$ per cent
below the 
prediction of \citet{evrard96}.  On the other hand, using \sax\ data,
\citet{ettori02} found a normalisation consistent with the predictions
(although the errors were large).

These studies had to rely largely on extrapolation to derive the
virial mass, and were limited by the low resolution and statistical
quality of the temperature profiles. With \xmm\ and \chandra\ we can
now measure the mass profile of clusters with unprecedented accuracy.
Using \chandra\ observations, \citet{allen01} derived an \MT\ relation
slope of $1.51\pm0.27$, consistent with the self-similar model, and
confirmed the offset in normalisation. However, their sample comprised
only 5 hot (i.e massive) clusters ($kT>5.5$~keV), and, due to the
relatively small \chandra\ field of view, their \MT\ relation was
established at $R_{2500}$ (i.e. about $\sim 0.3\rv$).

In a recent paper \citep*[hereafter, Paper I]{pointecouteau05}, we
measured the integrated mass profiles of ten relaxed, nearby clusters
observed with \xmm. The sample has an excellent temperature coverage,
from 2 to 9 keV. The mass profiles cover a wide range of radii (from
$0.01\rv$ to $0.7\rv$), and are particularly well constrained between
$0.1\rv$ and $0.5\rv$. In Paper I, we studied the structural
properties of the mass profiles, in order to test current scenarios
for the Dark Matter clustering. In this paper, these data are used to
establish a precise \MT\ relation up to the virial radius. In
Sect.~\ref{sec:data}, we describe how we derive the temperature and
mass data. In Sect.~\ref{sec:mt} we present and compare the
\MT\ relations at various density contrasts.  We discuss the reliability
of the X--ray mass estimates in Sect.~\ref{sec:mrelia} . The derived
\MT\ relations are discussed with respect to pre-\chandra/\xmm\ results
in Sect.~\ref{sec:dissobs}, and with expectations from models in
Sect.~\ref{sec:dissmod}. Our conclusions are presented in
Sect.~\ref{sec:concl}.

Throughout the paper, results are given for the currently-favoured
$\Lambda$CDM cosmology, with $H_0=70$~km/s/Mpc, $\Omega_m=0.3$ and
$\Omega_\Lambda=0.7$.
%=======================================================================

\section{The data \label{sec:data}}

\subsection{The mass at various density contrasts}
\label{sec:mass}

We used the mass profiles determined in Paper I to estimate the mass
at four density contrasts: $\delta=2500,1000,500$ and
$200$\footnote{We chose these $\delta$s for the following
reasons: $\delta=2500$ allows a direct comparison with previous
\chandra\ results; $\delta=1000$ is the density contrast limit of our
observations; $\delta=500$ corresponds to the edge of the virialized
part of clusters in a conservative approach; $\delta= 200$ is the
classical `virial' radius in an $\Omega=1$ universe.}.  We recall that
the mass profiles were derived from the observed density and
temperature profiles (corrected for PSF and projection effects) using
the hydrostatic equilibrium equation.  The mass and errors at each
radius of the temperature profile were calculated using a Monte Carlo
method.

In Paper I, we found that the mass profiles are well described by an
NFW-type model \citep{navarro97}:
\begin{equation}
M_{\rm NFW}(r) = \mv \frac{\ln(1+c_{200} x) -
c_{200}x/(1+c_{200}x)}{\ln(1+c_{200}) - c_{200}/(1+c_{200})}
\label{eq:nfw}
\end{equation}
where $x=r/\rv$. The concentration $c_{200}$, and the mass $\mv$, are
free parameters determined by the fitting procedure. $\rv$ is related
to $\mv$ via $\mv/\rv^{3} = (4\pi/3) 200 \rhoc(z)$. The best fitting
$\mv$ and $\rv$ values are recalled in Table~\ref{tab:mdel}.  The  
concentration parameters are given in Paper I. The corresponding NFW  
mass , $M_{\delta}$,  at any density contrast $\delta$, can be  
derived from the best fitting $c_{200}$ and $\mv$ values using  
Eq.~\ref{eq:nfw} ($M_{\delta}= M_{\rm NFW}(r_{\delta})$ with   
$M_{\delta}/r_{\delta}^{3}= (4\pi/3) \delta \rhoc(z)$). However, the  
computation of the error is not straightforward since the
uncertainties in  
$c_{200}$ and $\mv$ are correlated. To avoid this problem, for each  
$\delta$ under consideration, we refitted the observed mass profile
data with an NFW model using   
$M_{\delta}$ and the corresponding concentration as free parameters in  
the fit. The errors on $M_{\delta}$ can then be derived from  
standard $\Delta\chi^{2}$ criteria for one interesting parameter,  the  
concentration being optimised (minimum $\chi^{2}$) for any given value  
of  $M_{\delta}$. 
The resulting $M_{2500}$, $M_{1000}$ and
$M_{500}$ values are also shown in Table~\ref{tab:mdel}. We also list
the outermost radius (in units of $\rv$), and the corresponding
density contrast $\delta_{\rm obs}$, reached by the temperature
profiles (and thus the outermost extent of the measured mass profiles).

We will use these mass estimates to study the \MdT\ relation at
various density contrasts.  Our study is thus based on a parametric
model of the observed mass profiles, rather than directly on the
measured mass data.  Let us discuss this point in more detail. All of
the clusters are observed down to at least $\delta=1000$ ($R_{1000} =
(0.47\pm0.02)~\rv$, averaged over the whole sample), the only
exceptions being A1983 and MKW9 ($\delta_{\rm obs} \sim 1400$).  At
$\delta=1000$ and $\delta=2500$, using the best fitting model rather
than the data is simply equivalent to 'smoothing' the data (without
data extrapolation). 
We checked that this does not introduce a bias in the following way:
for each cluster, we estimated the mass at $\delta=2500$ by
interpolating the observed profile expressed as a function of density
contrast\footnote{$M(\delta)$ is readily derived from $M(r)$, using
$\delta= 3M(r)/4\pi\rhoc(z)r^3$ at each discrete radial value of
the mass profile.} in the log-log plane. We then compared the
interpolated value to that derived from the NFW fit to the mass
profile. In all cases the values are consistent within their $1\sigma$
errors. The ratio of the two values has a median value of $0.99$
across the sample, and there is no significant correlation with mass.
This reflects the fact that the NFW model is a good fit to these data,
particularly in the $0.1\rv-0.5\rv$ range (see Paper I).

As explained above, the estimates of $M_{2500}$ and $M_{1000}$ are
made (almost) without data extrapolation. However, the mass estimates at
$\delta=500$ and $\delta=200$ do involve extrapolation of the data.
The $M_{500}$ and $M_{200}$ estimates rely on the assumption that  
the
best fitting NFW model remains a good representation of the cluster mass
profile beyond $\delta_{\rm obs}$.  We further discuss the reliability
of this assumption in Sect.~\ref{sec:mrelia}.

\begin{table*}[t]
     \caption[]{Physical cluster parameters.  Masses are in units of
$10^{14}\, M_\odot$, and are given for a $\Lambda$CDM cosmology  with
$\Omega_{\rm m}=0.3$, $\Omega_{\Lambda}=0.7$, $H_0=70$~km/s/Mpc. Errors
are $1\sigma$ errors.}
     \label{tab:mdel}
     \begin{center}
%    $$
    \begin{tabular}{lccccccccc}
    \hline
    \hline
    Cluster & $z$ & $\kT~(\keV)$& $\rv $ (kpc)& $M_{2500}$& $M_{1000}$&
$M_{500}$& $\mv$ & $R_{obs}/R_{200}$ & $\delta_{obs}$ \\
    \hline
        A1983  &$0.0442$& $2.18\pm0.09$ &  $1103\pm 136$  & $  0.43\pm
0.09$  & $0.77\pm0.22$  & $1.09\pm0.37$ & $ 1.59\pm 0.61$   & 0.38 &
1455 \\
         MKW9  &$0.0382$& $2.43\pm0.24$ &  $1006\pm  84$  & $   
0.41\pm
0.07$  & $0.66\pm0.14$  & $0.88\pm0.20$ & $ 1.20\pm 0.30$   & 0.41 &
1401 \\
        A2717  &$ 0.0498$& $2.56\pm0.06$ & $1096\pm  44$  & $   
0.45\pm
0.04$  & $0.79\pm0.08$  & $1.10\pm0.12$ & $ 1.57\pm 0.19$   &  0.54 &
727 \\
        A1991  &$0.0586$& $2.71\pm0.07$ & $1106\pm  41$  & $  0.58\pm
0.05$  & $0.91\pm0.09$  & $1.20\pm0.12$ & $ 1.63\pm 0.18$   &  0.60 &
655 \\
        A2597  &$0.0852$& $3.67\pm0.09$ & $1344\pm  49$  & $  1.08\pm
0.07$ & $1.69\pm0.14$  & $2.22\pm0.22$ & $ 3.00\pm 0.33$   &  0.57 &
713  \\
        A1068  &$0.1375$& $4.67\pm0.11$ & $1635\pm  47$  & $  1.47\pm
0.07$  & $2.69\pm0.16$  & $3.87\pm0.28$ & $ 5.68\pm 0.49$   &  0.58 &
622 \\
        A1413  &$0.1430$& $6.62\pm0.14$ & $1707\pm  57$  & $  2.33\pm
0.13$ & $3.66\pm0.27$  & $4.82\pm0.42$ & $ 6.50\pm 0.65$   &  0.79 &
339 \\
         A478  &$0.0881$& $7.05\pm0.12$ & $2060\pm 112$  & $  3.12\pm
0.31$  & $5.43\pm0.70$  & $7.57\pm1.11$ & $ 10.8\pm 1.8$    &  0.58 &
650\\
      \pks &$0.1028$& $7.97\pm0.28$ & $1999\pm  77$  & $  3.32\pm 0.23$ &
$5.41\pm0.49$  & $7.27\pm0.75$ & $ 10.0\pm 1.2$    &  0.57 & 694 \\
       A2204  &$0.1523$& $8.26\pm0.22$ & $2075\pm  77$  & $ 3.62\pm 0.22$
& $6.11\pm0.51$  & $8.39\pm0.81$ & $ 11.8\pm 1.3$    &  0.61 & 580 \\
    \hline
    \hline
    \end{tabular}
\end{center}
Columns: (1) Cluster name; (2) Redshift; (3) Spectroscopic
temperature of the $0.1<r<0.5~\rv$ region; (4) $\rv$, roughly the
virial radius in numerical simulations;
(5,6,7,8) Total mass at density contrast $\delta=2500, 1000, 500, 200$,
derived from an NFW fit to the observed mass profile; (9)
Fraction of $\rv$ spectroscopically observed (outer radius
of the final temperature bin); (10) Corresponding density contrast.
\end{table*}

%%%%%%%%%%%%%%%%%%%%%%%%%%%%%%%%%%%%%
\begin{figure}
\begin{center}
\includegraphics[width=\columnwidth]{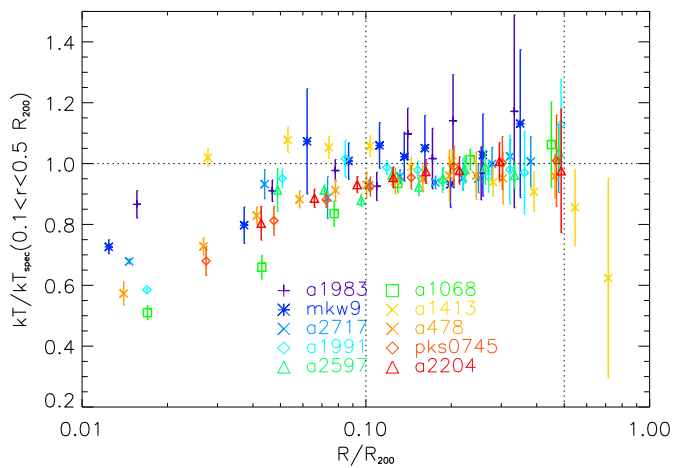}
\caption{Temperature profiles. The temperatures have been normalised
to the spectroscopic temperature measured in $0.1<r<0.5$~R$_{200}$
region; the radius has been scaled to  $\rv$. The profiles have been
corrected for PSF and projection effects (see Paper I for details).}
    \label{fig:ktprof}
\end{center}
\end{figure}
%%%%%%%%%%%%%%%%%%%%%%%%%%%%%%%%%%%%%

\subsection{Overall temperature \label{sec:ovkt}}

To investigate the $M-T$ relation, we need to define a global
temperature. For this quantity, we used the overall spectroscopic
temperature of the $0.1\rv\leq r \leq0.5\rv$ region. The lower
boundary of $0.1~\rv$ was chosen so as to avoid most of the cooling
core, where a large dispersion is observed in the temperature profiles
(Fig.~\ref{fig:ktprof}). The upper boundary is limited by the quality
of the spectroscopic data.  An upper boundary of $0.5~\rv$ appeared a
good compromise. Only the data from A1983 and MKW9 do not quite reach
this radius; they are however detected up to $\sim0.4~\rv$ (see
Table~\ref{tab:mdel}). Note that $0.5~\rv$ corresponds roughly to
$\delta=1000$.

For each cluster, we performed an isothermal fit of the spectrum
extracted within the $[0.1-0.5]\rv$ range, $\rv$ being derived from
the best fitting NFW model (see above). In the fit, the abundance was
let free and the $\nh$ was fixed to the 21 cm value (except for A478,
see \citealt{pointecouteau04}). We corrected the derived value for PSF  
blurring and
projection effects using the ratio of the mean emission-measure
weighted value of the temperature profile in the $[0.1-0.5]~\rv$
region after PSF/projection correction to the mean value before
correction (see Paper I for details on the correction procedure). The
correction factor is generally negligible and is always less than
5 per cent. The resulting temperature values are given in
Table~\ref{tab:mdel}.

We could have estimated a `mass-weighted' temperature in the $0.1\rv<
r < 0.5\rv$ region from the temperature profile. However, this
temperature would still be a `spectroscopic' temperature since it
would be derived from averaging over measured X-ray temperatures. It
would not, strictly speaking, be equivalent to the `mass-weighted'
temperature derived from numerical simulations. We thus preferred to
use the overall spectroscopic temperature, which is a directly
measured quantity and can also easily be estimated in numerical
simulations. Since the region is defined in scaled radius, it can be
derived from the simulated temperature profiles, using for instance
the approach proposed recently by \citet{mazzotta04}. In addition, we
note that only a global spectroscopic temperature can be usually
estimated for high $z$ clusters. In our approach, the extraction
region can be similarly defined and our definition thus allows a
consistent study of the evolution of the \MT\ relation.

We note that \citet{allen01} use a mass-weighted temperature,
$T_{2500}$, estimated from the temperature profile in the $r<R_{2500}
\sim 0.3\rv$ region. In practice, their definition is equivalent to
ours because i) the temperature profiles are fairly flat beyond the
cooling core region ($r> 0.1\rv$) in both studies, and ii) the cooling
core does not contribute much in mass to the average.  This can be
checked from Fig. 1 of \citet{allen01}, where $T_{2500}$ cannot be
distinguished from the spectroscopic temperature of the region beyond
the cooling core.

%=======================================================================

%=======================================================================
%%%%%%%%%%%%%%%%%%%%%%%%%%%%%%%%%%%%%

  \begin{table*}[]
    \caption[]{Results of power law fits to the \MdT\ and
    $R_{\delta}$--$T$ relation at various density contrasts
    $\delta$. The data are fitted with a power law of the form $h(z)
    M_{\delta} = A_{\delta} \times (\kT/5~\keV)^\alpha$ and $h(z)
    R_{\delta} = B_{\delta}\times (\kT/5~\keV)^\beta$, where $\kT$ is
    the overall spectroscopic temperature of the $[0.1\rv-0.5\rv]$
    region.  A $\Lambda$CDM cosmology is assumed: $\Omega_{\rm
    m}=0.3$, $\Omega_{\Lambda}=0.7$, and $H_0=70$~km/s/Mpc. }
    \label{tab:mt}
    \begin{center}
%    $$
   \begin{tabular}{rcccccccccccc}
   \hline
   \hline
\multicolumn{2}{l}{}  & \multicolumn{6}{c}{\MdT\ relation}&&
\multicolumn{2}{c}{$R_{\delta}$--$T$  relation}\\
\cline{3-8} \cline{10-12}
$\delta$&&
$A_{\delta}$ {\footnotesize ($10^{14}\,\msol$)}&  $\alpha$ &   
$\sigma_{log,raw}$ &  $\sigma_{log,int}$ & $\chi^2$(dof)&  nhp
&& $B_{\delta}$ (kpc)& $\beta$ \\
   \hline
\multicolumn{2}{l}{Whole sample}\\
\cline{1-2}
    200&&
     5.34$\pm$     0.22&     1.72$\pm$     0.10& 0.077 & 0.051  &  
14.49(8)&    0.07
    && 1674$\pm$23&  0.57$\pm$0.02 \\
    500&&
    3.84$\pm$     0.14&     1.71$\pm$     0.09&  0.064&  0.039  &   
12.65(8)&    0.12
    && 1104$\pm$13&  0.57$\pm$0.02 \\
   1000
   &&    2.82$\pm$     0.09&     1.71$\pm$     0.08&   0.053 &  0.027&   
10.74(8)&    0.22
   && 791$\pm$8&  0.57$\pm$0.02 \\
   2500
   &&    1.69$\pm$     0.05&     1.70$\pm$     0.07& 0.041   &  0.016&    
9.33(8)&    0.32
   && 491$\pm$4 &  0.56$\pm$0.02& \\
    \hline
\multicolumn{2}{l}{$T>3.5~\keV$}\\
   \cline{1-2}
   200&&    5.74$\pm$     0.30&     1.49$\pm$     0.17&  0.081 & 0.064 &  
    10.45(4)&    0.03
   && 1714$\pm$30&  0.50$\pm$0.05\\
    500&&    4.10$\pm$     0.19&     1.49$\pm$     0.15& 0.064   & 0.046  
&  8.34(4)&    0.08
    && 1129$\pm$17&
0.50$\pm$0.05\\
   1000&&    3.00$\pm$     0.12&     1.49$\pm$     0.14&  0.048 & 0.027  
&    5.91(4)&    0.21
   && 807$\pm$10&  0.50$\pm$0.04&\\
   2500&&    1.79$\pm$     0.06&     1.51$\pm$     0.11& 0.025 & -  &    
2.50(4)&    0.65
   && 500$\pm$5&  0.50$\pm$0.03&\\
\hline
\hline
   \end{tabular}
%$$
\end{center}
\footnotesize{
    Columns: (1) Density contrast $\delta$; (2, 3) Intercept and slope
for the \MdT\ relation: $h(z) M_{\delta} = A_{\delta}\times
(\kT/5\keV)^\alpha$ with standard errors; (4, 5) Raw and intrinsic  
scatter about the  best fitting relations in the log-log plane (see  
Sec.~\ref{sec:mt200}); (6)
Chi-squared and degree of freedom; (7) Null hypothesis probability
associated with the best fit; (8,9) Intercept and slope for the  
$R_{\delta}-T$
relation: $h(z) R_{\delta} = B_{\delta}\times (\kT/5\keV)^\beta$; }
\end{table*}

%=======================================================================

%=======================================================================

\section{The $M-T$ relation}
\label{sec:mt}
   \subsection{The $M_{2500}-T$ relation}
   \label{sec:mt2500}

In order to check the consistency of our \xmm\ results with the
\chandra\ study of \citet{allen01}, we first investigated the
$M_{2500}-T$ relation. For our sample, $\delta=2500$ corresponds to an
average radius of $[0.29\pm 0.02]~\rv$, where the mass is particularly
well constrained for all clusters.

The sample studied by \citet{allen01} comprises hot lensing clusters
($5.5$ to $15~\keV$). We thus considered only the sub-sample of
clusters with moderate to high temperatures (i.e. $T>3.5~\keV$), and
fitted the $M_{2500}$--$T$ relation using a power law model of the
form:
\begin{equation}
h(z) M_{\delta} = A_{\delta} ~ \left[\frac{\kT}{5~\keV}\right]^{\alpha}.
\label{eq:mt}
\end{equation}
Here and in the following, the fit is performed using linear
regression in the $\log$--$ \log$ plane, and the goodness of fit is
calculated using a $\chi^{2}$ estimator taking into account the errors
on both mass and temperature.  We used the routine FITEXY from  
numerical recipes \citep{numrec}.  Note that, as in the study of
\citet{allen01}, the masses are scaled by
$h(z)$, which corrects for the evolution expected in the standard
self-similar model. This scaling factor is small in our $z$ range
($h(z)=1.07$ at $z=0.15$) but varies between $\sim 1.05$ and $\sim
1.28$ for the \chandra\ clusters located at higher redshifts
($0.1<z<0.46$).

%%%%%%%%%%%%%%%%%%%%%%%%%%%%%%%%%%%%%
\begin{figure}[]
\begin{center}
   \includegraphics[width=\columnwidth]{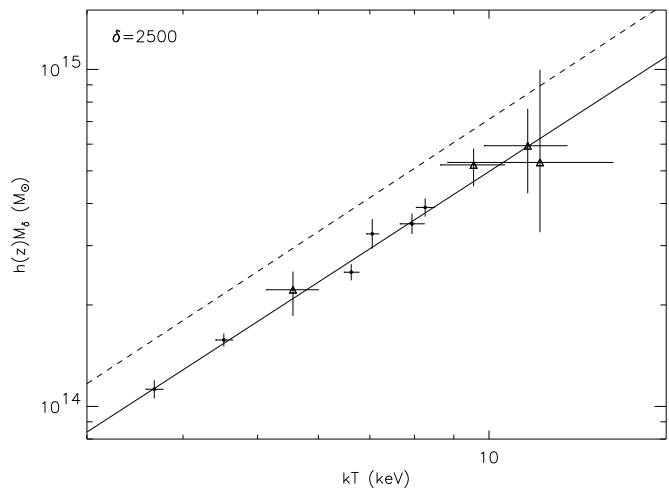}
   \caption{ The $M-T$ relation at $\delta=2500$ as seen by \xmm\ from
   the observation of 6 hot ($\kT>3.5$~keV), relaxed clusters. Filled
   squares show the \xmm\ data points; the full line shows the best
   fitting power law. The data on 4 published Chandra clusters
   (triangles) have been added to the fit but due to their larger
   uncertainties, they do not change the parameters of the fit to the
   \xmm\ data 
   only (see text).  The dashed black line is the prediction from
   adiabatic numerical simulations \citep{evrard96}.
}
   \label{fig:mt2500}
\end{center}
   \end{figure}

%%%%%%%%%%%%%%%%%%%%%%%%%%%%%%%%%%%%%
    \begin{figure*}[t]
\begin{center}
   \includegraphics[width=\textwidth]{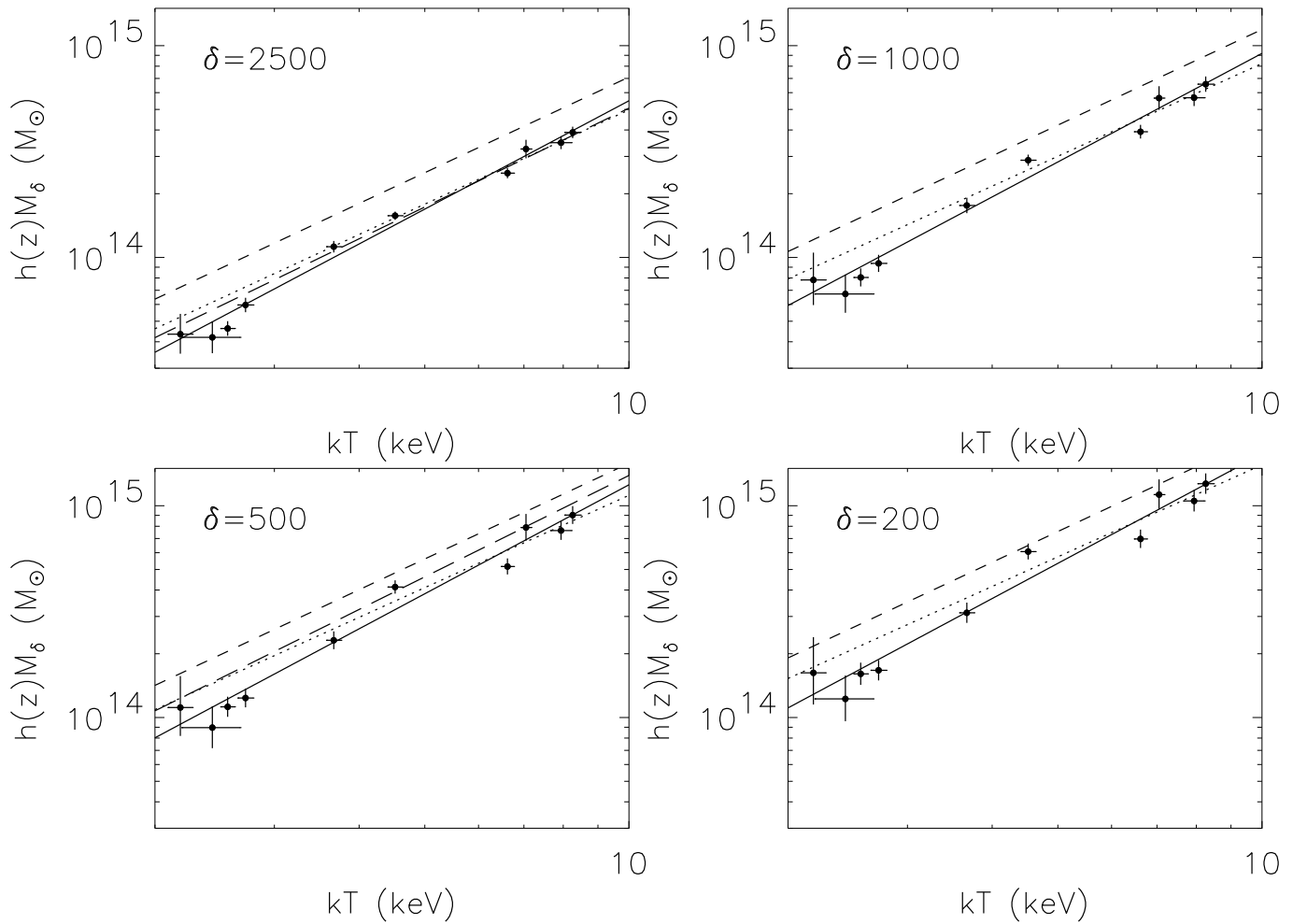}%
   \end{center}
\caption{ The $M-T$ relation as seen by \xmm\ from a sample of 10
clusters covering a temperature range from 2 to 9~keV. From top to
bottom and left to right, the $M-T$ relation is given at the density
contrasts $\delta$ of 200, 500, 1000 and 2500 with respect to the
critical density of the Universe. Measurements are plotted with error
bars. In each panel, the best fit for the whole sample is overplotted
as a solid line, and the best fit for the hot cluster subsample is
plotted as a dotted line.  The predicted relation from adiabatic
numerical simulations \citep{evrard96} is overplotted as a dashed
line. The long-dashed line (panels $\delta=500$ and $\delta=2500$) is
the relation derived from a numerical simulation including radiative
cooling, star formation and SN feedback \citep{borgani04}.
    }
   \label{fig:mt}
   \end{figure*}
%%%%%%%%%%%%%%%%%%%%%%%%%%%%%%%%%%%%%

The data are well fitted by a power law ($\chi^2/\textrm{dof}=2.5/4$).
The slope, $\alpha = 1.51\pm0.11$, is perfectly consistent with the
expectation from the standard self-similar model, and with the results
from \chandra\ observations ($\alpha=1.51\pm0.27$).  The derived
normalisation, $A = (1.79\pm0.06) \times 10^{14}\ \msol$, is also
consistent with the \chandra\ normalisation (see
Table~\ref{tab:comp}).  As noted by \citet{allen01}, such a
normalisation is discrepant with the value derived from numerical
simulations including gravitational heating only: our measured value
is about $\sim30$ per cent below the prediction of
\citet{evrard96}. When the \chandra\ data for 4 of the 5  clusters
studied by \citet{allen01} are added to the present data
set\footnote{The fifth cluster is \pks, which is common to both
samples. We use only the \xmm\ measurement here.}, the
best fitting values are almost unchanged ($\alpha=1.52\pm0.1$ with the
same intercept). This is due to the larger uncertainties in the
\chandra\ temperature and mass determinations compared to those measured
here (see Fig.~\ref{fig:mt2500}).  Figure~\ref{fig:mt2500} shows the
best fit for the combined \xmm\ and \chandra\ data compared to the
expectations from the adiabatic numerical simulations of
\citet{evrard96}.

Still working at $\delta=2500$, we performed a fit over the whole
\xmm\ sample, i.e. now including the four low mass systems. We obtain
$\alpha=1.70\pm 0.07$, and a normalisation $A=(1.79\pm0.06)\times
10^{14}~\msol$. The fit is acceptable, although formally less good
($\chi^2/\textrm{dof}=9.33/8$). The slope now differs significantly
from the expected value of $\alpha=1.5$, and is just barely consistent
with it at a $3\sigma$ level.  This is further discussed in
Sec.~\ref{sec:dissmodslope}.

%=======================================================================

   \subsection{The \MdT\ relations up to the virial radius
\label{sec:mt200}}

Figure~\ref{fig:mt} shows the \MdT\ relations at various $\delta$,
together with the best fitting power law (Eq.~\ref{eq:mt}) in each
case, and the prediction from the numerical simulations of
\citet{evrard96}. The best fitting slopes and normalisations are listed
in Table~\ref{tab:mt},  together with the standard errors. The  
best fits are listed and plotted both for
the whole sample, and for the sub-sample of hot clusters. The  
corresponding $R_{\delta}$--$T$ relations are also given in the Table.

The normalisation and slope are nearly independent parameters for  
the whole sample.  The covariance in $\log(A_{\delta})$ and $\alpha$,  
normalised to the product of their  standard errors, is small:   
$0.045, 0.071, 0.092$ and $0.11$ for $\delta=2500,1000,500$ and $200$,  
respectively.  This is due to our choice of the pivot of the \MdT\   
relation, $\kT = 5~\keV$ (Eq.~\ref{eq:mt}), close to the mean  
temperature of the whole sample ($4.8 \keV$) or the median value of its  
temperature range ($5.2 \keV$). The normalisation for other  
pivots can be derived using our best fitting values from Eq.~\ref{eq:mt}.  
For instance, for a pivot  at $6 ~\keV$, used in several works on 
cluster scaling relations, the normalisation $A_{\delta}$ is $\sim  
36$ per cent higher. The relative error, $\sigma_{A_{\delta}}/A_{\delta}$, is  
increased by $\sim 10$ per cent.

  We have also computed the raw and intrinsic scatter about the best fitting  
relations in the log-log plane. They are given in Table~\ref{tab:mt}.  
To estimate the raw scatter, we used  the orthogonal distances to the  
regression line, weighted by the  error \footnote{For a linear relation  
of the form $Y= a X + b $, and a sample of N data points  
($Y_{i},X_{i}$) with errors $\sigma_{Y_{i}}$ and $\sigma_{X_{i}}$,  an  
estimate of the square of the raw scatter is: $\sigma_{raw} ^{2}=  
\frac{1}{N-2} \sum_{i=1}^{N} w_{i} (Y_{i} - a X_{i} -b)^{2}$, where    
$w_{i} =  \frac{1/\sigma_{i}^{2}} {(1/N) \sum_{i=1}^{N}  
1/\sigma_{i}^{2}}$ with $\sigma_{i}^{2} = \sigma_{Y_{i}}^{2} + a^{2}  
\sigma_{X_{i}}^{2}$. Here  $Y = \log(M_{\delta}), X = \log(T)$. }. The  
intrinsic scatter is computed from the quadratic difference between the  
raw scatter and the scatter expected from the statistical errors.

The behaviour of the $M_{2500}$--$T$ relation is reproduced at all
other density contrasts.  The slope is stable on all spatial scales:
the variation is at most $10$ per cent of the statistical error. It is
always 
consistent with the expected $\alpha=1.5$ value for the sub-sample of
$T>3.5~\keV$ clusters, whereas it steepens to $\alpha=1.7$ when the
cool clusters are included. Similarly, the normalisation remains $\sim
30$ per cent below the value from adiabatic numerical simulations of
\citet{evrard96} at all $\delta$.

This stable behaviour is a direct consequence of the self-similarity of
the mass profiles (Paper I). For an NFW type profile, the ratio of the
masses at different density contrasts only depends on the
concentration parameter, and $M_{\delta} = M_{2500} F(c,\delta)$.  If
all clusters had exactly the same concentration parameter (i.e.,
perfect self-similarity) the \MdT\ relations at various $\delta$
should differ only in their normalisation: $M_{\delta} = A(2500)
F(c,\delta) T^{\alpha}$, for $M_{2500} = A(2500)T^{\alpha}$.  The
clusters in our sample are not perfectly self-similar but there is no
significant variation of $c$ with mass, and thus temperature (see
Fig. 3 of Paper I). This explains the observed invariance of the
slope.  Furthermore, the observed concentration parameter is
consistent with theoretical expectations (Paper I). As a consequence,
the variation with $\delta$ of the normalisation, $A(\delta)= A(2500)
F(c,\delta)$, follows expectations, and the offset with respect to  
simulations observed at
$\delta=2500$ remains the same at all $\delta$ (see also
Fig.~\ref{fig:Acomp}, and Fig 13 of \citealt{pratt02}).

However, the quality of the power law fit decreases with decreasing
$\delta$ (see Table~\ref{tab:mt} and also Fig.~\ref{fig:mt}).  The
reduced $\chi^2$ increases and the corresponding null hypothesis
probability for the whole sample varies from $0.32$ at $\delta=2500$
to $0.07$ at $\delta=200$. This behaviour corresponds to an increase
of the intrinsic scatter in the observed \MdT\ relation (see
Table~\ref{tab:mt}). For the whole sample, the worst fit and largest
intrinsic scatter is observed at $\delta=200$.  The regression method
we used (Sect.~\ref{sec:mass}) is strictly valid only if the intrinsic
scatter is negligible as compared to the statistical scatter.  This is
not always the case (see Table~\ref{tab:mt}), and we first checked if
our results could be affected, using the $M_{200}$--$T$ relation for
the whole sample (i.e., the worst case). We refitted the data using
the orthogonal BCES method \citep{akritas96}. While this is the
least-biased regression method when both measurement errors and
intrinsic scatter are present, it is less accurate that the $\chi^2$
method when the intrinsic scatter is negligible. The best fitting values remain unchanged (within $0.5$ per cent) and the standard error
estimates are only slightly larger (by $\sim 15$ per cent).

The regression method we used is thus justified.  Nevertheless, the
derived intrinsic scatter should not be over-interpreted. The cluster
sample is small and is certainly not representative of the entire
cluster population. In particular, it is heavily biased towards the
more relaxed clusters. Moreover, the increased intrinsic scatter at
low $\delta$ may be an artifact of the method we used to derive the
various $M_{\delta}$. These were derived from an NFW fit to the observed
mass profile, using $M_{\delta}$ as a free parameter (see
Sect.~\ref{sec:mass}).  In the NFW fit, $M_{2500}$ is extremely
well constrained by the data around $\delta=2500$, quasi-independently
of the shape of the observed mass profile.  On the other hand,
$M_{\delta}$ at low $\delta$, beyond the maximum radius of
observation, can be viewed as an `extrapolation' of the NFW model best
fitting the observed mass profile. It thus depends both on the
normalisation of the observed profile (basically $M_{2500}$) and on
its shape. The shape parameter (the concentration $c$) is very
sensitive to the data at small radii, in particular in the cooling core
region, where the mass profile is least well constrained, and where
there could be systematic errors due to the PSF/projection correction.
As a result we expect, as observed, increasing statistical errors on
$M_{\delta}$ as $\delta$ decreases, and a corresponding increase of
the raw scatter in the \MdT\ relation (reflecting the scatter in $c$,
see Fig. 3 of Paper I). The increase might be larger than that
expected purely from the increase of statistical errors due to
intrinsic scatter in $c$ and/or systematic errors on $c$.

%=======================================================================

\section{ Reliability of  X-ray mass estimates}
\label{sec:mrelia}

As discussed in Paper I, there is an excellent {\it quantitative}
agreement in {\it shape} between the X-ray mass profiles used in this
work and the profile predicted by numerical simulations. The observed
scaled profiles are well-described by the quasi-universal cusped
profile (NFW-type) now found in all CDM simulations, and have
concentration parameters as expected for their mass.  As concluded in
Paper I, this suggests that the Dark Matter collapse is well
understood, at least down to the cluster scale. In turn, this gives us
confidence in the \xmm\ mass estimates, not only in the observed
radial range, but also where we have extrapolated beyond it (i.e at
$\delta < \delta_{\rm obs}~\sim 1000$).  By using the best fitting NFW
model to estimate $M_{500}$ and $M_{200}$, we have implicitly assumed
that this model remains valid beyond $\delta_{\rm obs}$
(Sec.~\ref{sec:mass}). It would be surprising if this were not the case
since i) it is consistent with the theoretical predictions above
$\delta_{\rm obs}$, and ii) in one case (A1413), we were even able to
check the validity of the NFW profile down to $\delta<500$.

Strictly speaking, the good agreement  between the observed and
predicted shape of the mass profiles does not mean that the {\it
absolute} value of the X-ray mass is correct. It could be subject to
systematic errors. However, for the correct universal shape of the mass
profile to be recovered, this systematic error would have to be the same,
within the statistical errors at all observed $\delta$, whatever the
cluster temperature.

One possible source of such systematic error is a departure from
hydrostatic equilibrium (HE).  The recent simulations of \citet{kay04a}
suggest that the mass determined from the HE equation underestimates
the true mass, due to residual gas motion.  The effect is about the
same at all radii up to $\delta=500$. It is of the order of $15$ per
cent for adiabatic models and of $10$ per cent for models including cooling and
feedback, with typical variations of $\pm5$ per cent.  Such variations would
not significantly change the shape of the X-ray mass profiles, taking
into acccount our statistical errors. Thus the measured $M_{2500}$ and
$M_{1000}$ values, and thus the corresponding normalisation of
the$M_{2500}$--$T$ and $M_{1000}$--$T$ relations, could well be $\sim
10-15$ per cent too low. The offset would be the same for the
$M_{500}$--$T$ and $M_{200}$--$T$ relations, since they are derived
from `extrapolation' of the NFW model. Note that this is probably an
upper limit, since we focus on particularly relaxed clusters.

Another possible source of systematic error is that associated with
errors in estimated temperatures from uncertainties in the  
instrument calibration. Possible errors are of the order of
$10$ per cent. This value is consistent with the systematic difference  
observed between temperature derived with \xmm\ and \chandra\, with the  
former being on average $0.92\pm0.08$ times the latter  
\citep{kotov05}.  Since the mass derived from the HE equation scales  
as $M\propto T$, error in $T$
would translate into a systematic error on the estimate of the
`real' mass of clusters. However that would not change the
normalisation of the `observed' \MT\ relation, since that depends only
on the {\it shape} of the temperature profile.

%%%%%%%%%%%%%%%%%%%%%%%%%%%%%%%%%%%%%%%%%%%%%%%%%%%%%%%%%%%%%%%%%

\section{Comparison with previous determinations}
\label{sec:dissobs}

In the present study, as discussed above, an NFW profile has been used
to describe the integrated mass profile, derived, assuming HE, from the
observed density and temperature profiles. A similar approach
\footnote{The method differs lightly from ours. \citet{allen01}  
predict the temperature profile corresponding to a given NFW mass  
profile and the observed surface brigthness profile and fit it to the  
observed temperature profile.}
was used
in the \chandra\ study of \citet{allen01}.
Previous \rosat/\asca\ studies also estimated the mass from the HE
equation, but assumed a $\beta$--model for the gas density profile and a
polytropic (or even isothermal) temperature profile \citep{horner99,
nevalainen00,finoguenov01, xu01, castillo03, sanderson03}.  Data were
of poorer spatial resolution and statistical quality, and had less radial
extent, thus requiring more extrapolation, particularly for low
mass clusters. This could introduce systematic errors and biases,
particularly at low $\delta$, since a mass profile derived from an
isothermal or polytropic $\beta$-model is not consistent with an NFW
profile at large radii \citep[see Fig 11 of ][]{neumann99}. On the
other hand, the latest \rosat/\asca\ studies of the \MT\ relation
\citep{finoguenov01,sanderson03}  are superior in terms of the size of  
the
cluster samples, and their wide and homogeneous coverage in
temperature.  Their results are compared to ours in
Table~\ref{tab:comp}.

\citet{finoguenov01} established the $M_{500}$--$T$ relation for 39
clusters with \asca\ temperature profiles.  Interestingly, their
results are consistent with ours within the uncertainties
(Table~\ref{tab:comp}). This suggests that systematic errors are not
dominant over statistical errors at $\delta=500$ in this \rosat/\asca\
study.  However, our normalisation is on the upper side of their
allowed values.
The slope they find for their hot cluster subsample ($T > 3~\keV$) is,
as we have found, consistent with the expected $\alpha=1.5$
value. When
\citet{finoguenov01} included all clusters (down to $T\sim 0.9~\keV$),
they found a steepening of the $M_{500}$--$T$ relation:
$\alpha=1.78\pm0.1$. The effect is larger than in our case: we find
$\alpha=1.71\pm0.09$ (although the difference is not significant).
However, our sample does not reach quite such low temperatures  
and the
difference could also reflect a progressive steepening of the \MT\
relation toward low masses.

%%%%%%%%%%%%%%%%%%%%%%%%%%%%%%%%%%%%%%%%%%%%%%%%%%%%%%%%%%%%%%%%%
\begin{table}[t]
     \caption[]{Comparison of the present results with $M-T$ relations
     from the literature. $\alpha$ is the logarithmic slope of the
     relation and A is the normalisation at $\kT=5~\keV$, in units of
     $10^{14}\,\msol$ for $H_0=70$~km/s/Mpc.
}
     \label{tab:comp}
     \begin{center}
\begin{tabular}{llll}
\hline
\hline
Reference$^{\mathrm{a}}$  & \multicolumn{1}{c}{$A $} &
\multicolumn{1}{c}{$\alpha$}&  Method$^{\mathrm{b}}$ \\
\hline
\bf{$\delta=2500$}\\
\cline{1-1}
Observation\\
\cline{1-1}
Present work&                $1.69\pm0.05$    &         $1.70\pm0.07$&
$M_{\rm NFW}$, $T>2.0~\keV$\\
Present work&                $1.79\pm0.06$      &       $1.51\pm0.11$&
$M_{\rm NFW}$, $T>3.5~\keV$\\
ASF01&         $1.88\pm0.34$$^{\mathrm{c}}$     &       $1.51\pm0.27$&
$M_{\rm NFW}$, $T>5.5~\keV$ \\
SPF03&    $1.2\pm0.4$$^{\mathrm{c}}$            &       $1.84\pm0.14$&   
   $M_{\rm
\beta\gamma}$, $T > 5.5~\keV$\\
\cline{1-1}
Theory\\
\cline{1-1}
EMN96&  $2.5$  & 1.5  &$\Tem$, adiabatic simul.\\
BMS04&      $1.73\pm0.35$ &$1.55\pm0.05$ &     $\Tm$,$T>2.0~\keV$\\
KSA04&         $1.97\pm0.03$ & $1.54\pm0.05$& $\Tm$\\
\hline
\bf{$\delta=500$}\\
\cline{1-1}
Observation\\
\cline{1-1}
Present work&           $3.84\pm0.14$&$1.71\pm 0.09$&   $M_{\rm NFW}$,
$T>2.0~\keV$\\
Present work&            $4.10\pm0.19$&$1.49\pm 0.15$&  $M_{\rm NFW}$,
$T>3.5~\keV$\\
FRB01     & $3.26\pm0.60$$^{\mathrm{c}}$                  &$1.48\pm  
0.11$
&   $M_{\rm \beta\gamma}$, $T>3~\keV$\\
FRB01  & $3.31\pm0.45$$^{\mathrm{c}}$                        &$1.78\pm
0.10$ &    $M_{\rm \beta\gamma}$, $T>0.9~\keV$\\
\cline{1-1}
Theory\\
\cline{1-1}
EMN96 &5.6 &1.5 &   $T_{\rm ew}$, adiabatic simul.\\
VBB02 & $3.6$ & $\sim 1.7$& $\Tem$ \\
BMS04&$4.6\pm0.2$&$1.59\pm0.05$& $\Tem$,$T>0.7~\keV$\\
RMB05&   $7.2 \pm 0.5$ &  $1.66\pm0.09$& $\Tsl$, $T>1~\keV$\\
RMB05&   $4.2 \pm 0.2$ &  $1.53\pm0.05$& $M_{\rm \beta\gamma}$, $\Tsl$;
$T>1~\keV$\\
\hline
{\bf $\delta=200$}\\
\cline{1-1}
Observation\\
\cline{1-1}
Present work&       $ 5.34\pm0.22$&$1.72\pm0.10$ & $M_{\rm NFW}$,  
$T>2~\keV$\\
Present work&      $ 5.74\pm0.30$&$1.49\pm0.17$ & $M_{\rm NFW}$,  
$T>~3.5\keV$\\
SPF03  & $4.5\pm0.3$$^{\mathrm{c}}$      & $1.84\pm0.06$& $M_{\rm
\beta\gamma}$, $T>0.6~\keV$\\
\cline{1-1}
Theory\\
\cline{1-1}
EMN96& 7.4 & 1.5 & $\Tem$, adiabatic simul.\\
MTK02&7.6&$1.61$&  $\Tem$,$T>2.0~\keV$ \\
\hline
\hline
\end{tabular}
\end{center}
\noindent Notes:  (a): References: (ASF01) \citet{allen01}; (BMS04)
\citet{borgani04}; (EMN96) \citet{evrard96}; (FRB01)
\citet{finoguenov01}; (KSA04) \citet{kay04b};  (MTK02)
\citet{muanwong02}; (PW) present work ;   (RMB05) \citet{rasia05};
(SPF03) \citet{sanderson03}; (VBB02) \citet{voit02}; (b): Method:
($M_{\rm NFW}$): mass estimated using an NFW model to describe the
mass profile; ($M_{\rm \beta\gamma}$): mass estimated using a
polytropic $\beta$-model for the gas distribution; ($\Tm$):
mass-weighted temperature; ($\Tem$): emission-weighted temperature;
($\Tsl$): spectroscopic-like temperature as defined in
\citet{mazzotta04}.
(c): The normalisation at $5~\keV$ is derived from the published
normalisation at $1~\keV$ and the best fitting slope. The fractional  
error
has been assumed to be the same, which is conservative.

\end{table}
%%%%%%%%%%%%%%%%%%%%%%%%%%%%%%%%%%%%%%%%

The same remark holds for the results of \citet{sanderson03}, who
derived a slope of $\alpha=1.84\pm0.06$ for the $M_{200}$--$T$
relation, which is barely consistent with our value. Their large
sample includes 66 clusters in the $[0.5-15]~\keV$ temperature range.
Furthermore, their normalisation at $\delta=200$ is significantly
lower than ours (by $15$ per cent, Table~\ref{tab:comp}). This may
reflect the introduction of systematic errors when extrapolating
polytropic models down to $\delta$ as low as $200$.  However, such
systematic errors could not explain the discrepancy between our
results and theirs at $\delta=2500$.  At that density contrast, their
slope for a hot cluster ($T>5.5~\keV$) sub-sample is
$\alpha=1.84\pm0.14$, a result which is inconsistent with ours at the
$\sim 95$ per cent confidence level.  Their normalisation at
$\delta=2500$, $(1.2\pm0.4) \times 10^{14}~\msol$, is lower than our
value of $(1.69\pm 0.05) \times 10^{14}~\msol$ and is only barely
consistent with the \chandra\ results.

\citet{sanderson03} have suggested that the discrepancy (with respect
to the \chandra\ results) might be
related to the dynamical state of the clusters in the different
samples.  Both the \chandra\ study and the present \xmm\ study focus
on particularly relaxed clusters, which is not the case in
\citeauthor{sanderson03}'s study. However, we note that
\citeauthor{finoguenov01}'s sample does not discriminate in terms of
dynamical state, and their results are in good agreement with
ours. That said, the \MT\ relation could well depend on the exact
dynamical states of the clusters in the sample in question, an effect
which is not trivial to predict. The numerical simulations of
\citet{rowley04} show that clusters with substructure
tend to lie below the mean $T$-$M$ relation, probably due to
incomplete thermalisation (their Fig. 15). However, for the same
reason, we would 
expect the X-ray mass to underestimate the true mass
\citep{kay04a}, moving unrelaxed clusters back closer to the mean
relation. A \xmm\ study of the \MT\ relation for an unbiased sample of
clusters is needed to assess the effect of cluster dynamical state on
the measured \MT\ relation. The relation should ideally be compared to
numerical simulations and lensing mass data. In any case, we do not
confirm \citeauthor{sanderson03}'s results, at least for the relaxed
clusters considered here.

Finally, it is of interest to compare the present results with those
of \citet{ettori02}, who also use an NFW model to estimate
masses\footnote{In 
their approach, the predicted temperature profile was fitted to the
observed \sax\ temperature profile.}. Their \MT\ relation for a
relaxed sub-sample of 12 clusters with $T > 3~ \keV$ can be directly
compared with our relation for hot clusters. At $\delta=2500$, they
found $\alpha=1.88\pm0.27$, which is marginally consistent with our
value, and $\alpha=2.3\pm 0.4$ at $\delta=500$, a value clearly
rejected by our data. However, there is a large scatter in their \MT\
relation at $\delta=2500$, which becomes dramatic at $\delta=500$ 
(their Figure 5).  We do not observe such a scatter. It may reflect
systematic errors connected to the extrapolation of the NFW model. As
we discussed above in Sect.~\ref{sec:mt200}, the precision on
extrapolated mass depends on the precision on the concentration
parameter (i.e., the shape of the mass profile, especially in the  
center), which is more difficult to constrain with \sax\ than
with \xmm, in particular due to the larger \sax\ PSF.

\section{Comparison with theoretical predictions}
\label{sec:dissmod}

The temperature structure of the ICM in a cluster is the
result of the complex interplay between gravitational processes (i.e.,
the evolution of the gas in the Dark Matter potential), and of any
other process that can affect the gas entropy (e.g., radiative cooling and
heating from galaxy feedback). The theoretical \MT\ relation -- which
should be viewed rather as a $T$ versus $M$ relation when predicted
from theoretical studies -- depends on the exact modelling of all
these processes. Moreover, as we discuss below, comparison of
observations and theory also depends on the exact definition of the
`average' temperature, since the gas is never perfectly isothermal.

\subsection{The normalisation of the \MdT\ relation at $5~\keV$}

The normalisation of the \MT\ relation is particularly well
constrained by our study, the statistical error now being less than
$\sim 5$ per cent at $5~\keV$.  The value of the normalisation depends
on the (sub)sample considered because it is correlated with the slope.
However by choosing a reference temperature of $5~\keV$, close to the
median temperature, we minimize the effect, and the difference is of
the order of the statistical error.  We can thus first compare our
results with the predicted values, quasi-independently of the slope
issue.

%%%%%%%%%%%%%%%%%%%%%%%%%%%%%%%%%%%%%
\begin{figure}[]
\begin{center}
   \includegraphics[width=\columnwidth]{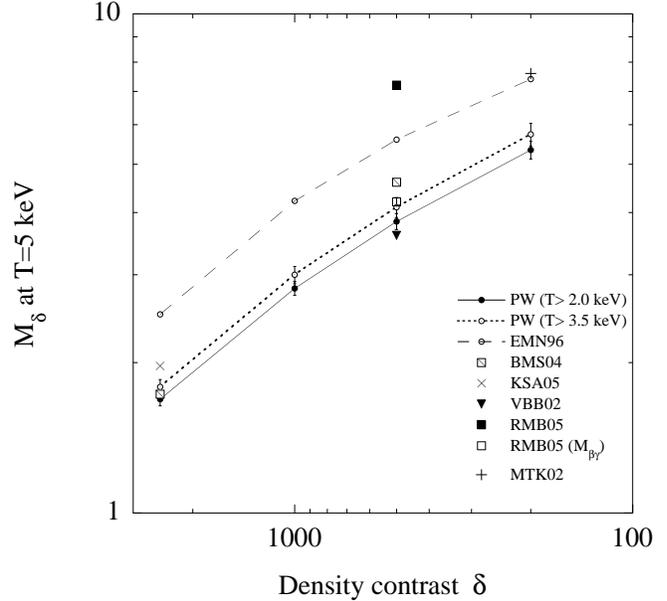}
   \caption{The normalisation of the \MdT\ relation at $T=5~\keV$ for
   various $\delta$. The results of theoretical works (see
   Table~\ref{tab:comp} for references) are compared to the values
   derived in the present work (PW)
.}
    \label{fig:Acomp}
\end{center}
   \end{figure}
%%%%%%%%%%%%%%%%%%%%%%%%%%%%%%%%%%%%%

\subsubsection{Adiabatic models}

The difference with the prediction from adiabatic models is not
dramatic, particularly when the dispersion among various adiabatic
simulations is taken into account. The normalisation is $30$ per cent
below the prediction of \citet{evrard96}, compared to typically more than
$50$ per cent difference in the normalisation derived from different
adiabatic simulations (Table~3; \citealt{henry04}).
However, as discussed by \citet{muanwong02}, higher resolution
simulations  tend to predict higher normalisation, thus exacerbating
the discrepancy with our results.
Adiabatic simulations probably fundamentally fail to predict the
correct normalisation of the \MT\ relation. The observed discrepancy
could in principle be due to incorrect modelling of the  dark  
matter
component itself, since it is this which drives the
potential. However, we have
already argued that this is not the case from the excellent agreement
between the observed and simulated mass profiles (Paper I and
Sec.~\ref{sec:mrelia}). A more likely explanation is that
non-gravitational processes affect the \MT\ relation, as they affect
all other scaling relations.

\subsubsection{Non-adiabatic models}

 From theoretical arguments and from observations of the gas entropy,
it is now clear that both cooling and galaxy feedback have to be taken
into account when discussing relations
involving the ICM \citep{voit03,pratt05}. We thus focus on published
theoretical studies that
include both phenomena.  Their predictions are compared with our
results in Table~\ref{tab:comp} and in Fig.~\ref{fig:Acomp}. All of
the quoted simulations are made in the `concordance' $\Lambda$CDM
cosmology, and use the same definition of $M_{\delta}$ as in the
present work. However, they do not always use the same definition
of the temperature (see below).

 At a given mass, cooling and galaxy feedback increase the gas  
entropy
as compared to the value attained by pure gravitational heating. 
In many scenarios, these processes increase the
temperature and, as expected, a lower normalisation is
found when these processes are included in a given numerical
simulation \citep{muanwong02,thomas02} or analytical model  
\citep{voit02}.
Recent non-adiabatic simulations seem to be quite successful at
reproducing the observed $M_{2500}$--$T$ relation. The normalisation
derived by \citet{borgani04}, for a sample of simulated clusters with
temperatures $T > 2$ keV, is in perfect agreement with our value (see  
also
Fig.~\ref{fig:mt}),
while the normalisation derived by \citet[see also
\citealt{thomas02}]{kay04b} is only $\sim 10$ per cent too high.  These works
used the mass weighted temperature, $T_{\rm m,2500}$, estimated within
$R_{2500}$, which is directly linked to the thermal energy. As
discussed in Sect.~\ref{sec:ovkt} our spectroscopic temperature, $\Ts$,
should be close to the mass weighted temperature in that region. Thus,
the good agreement between the observed $M_{2500}$--$\Ts$ and
predicted $M_{2500}$--$\Tm$ relations is encouraging, and suggests
that the thermal energy content in this central region is roughly
correctly modelled.

Up until recently, the standard temperature definition used to mimic
X-ray observations in numerical simulation studies was the
emission-weighted temperature $\Tem$. Using this temperature
definition at $\delta=500$, the normalisation of \citet{borgani04} is
now too high by about $12-20$ per cent, the normalisation found by
\citet{voit02} is marginally too low and the normalisation found by
\citet{muanwong02} at $\delta=200$ is $30$ per cent too high (
Fig.~\ref{fig:Acomp}).  It is likely that the variabality of these
results is linked to differences in the various physical models used.

Independent of the physics, a crucial point seems to be the exact
definition of the temperature. Recently, \citet{mazzotta04} introduced
the spectroscopic-like temperature ($\Tsl$) in order to better
reproduce the temperature obtained from spectral fits when the ICM is
multi-temperature. \citeauthor{mazzotta04} show that $\Tsl$ is biased
towards the lower values of the dominant thermal component, and that
in general $\Tem$ overestimates $\Tsl$.  Unfortunately, this
exacerbates the disagreement between observed and simulated
$M_{500}$--$T$ normalisations.  Using $\Tsl$, \citet{rasia05}
over-predict a normalisation, relative to our values, by a factor as
large as $\sim 1.8$.  Note that \citeauthor{rasia05} use the same
physical model as \citet{borgani04} and yet their normalisation is
$\sim 50$ per cent higher (see Table~\ref{tab:comp} and also their Fig 2).  
We note that the earlier work of \citet{mathiesen01}, based on  
spectroscopic temperatures of adiabatic numerical simulations, showed
a smaller effect.
Strictly speaking these temperatures were estimated with $R_{500}$,
whereas our spectroscopic temperature measurement $\Ts$ is interior to
$0.5\rv (\sim R_{1000}$).  A1413 is the only cluster for which we have
data up to $\delta=500$. The spectroscopic temperature within
$\delta=500$, $T_{\rm s,500}$ is only slightly smaller (by $3$ per cent) than
$\Ts$.  This would increase the normalisation of the $M_{500}$-$T$
relation by less than $5$ per cent if we used $T_{\rm s,500}$ (assuming the
same correction factor for all clusters).

It thus appears that there is a genuine disagreement between observed
and predicted normalisation of the $M_{500}$--$T$ relation. One
interpretation, as proposed by  \citet{rasia05}, is that the X-ray mass
underestimates the `true' mass \citep[see
also][]{borgani04,muanwong02}. Using $\Tsl$
temperatures, they estimated the value of $M_{500}$
that an X-ray observer would derive from their simulation using the HE
equation and a polytropic \betamodel. The resulting $M_{500}$ - $\Tsl$  
is indeed now in good
agreement with our observation (Table \ref{tab:comp} and
Fig.~\ref{fig:Acomp}). However, the normalisation is $4.2 \pm 0.2\times
10^{14}\ \msol$, as compared to $7.2 \pm 0.5\times 10^{14}\ \msol$ when
using the `true' theoretical mass.  This corresponds to a very serious
underestimate of the mass by X--ray observations: the `true'  $M_{500}$
   mass  of clusters  would be a factor $\sim 7.2/4.2= 1.7$ higher than
the X-ray mass. We think this is very unlikely, at least  for the
masses estimated as in the present work. Firstly, our approach -- fitting  
an NFW
model and extrapolating the mass profiles -- is more sophisticated than
the simple polytropic \betamodel\ approach.
Secondly, we note again the
excellent quantitative agreement of our mass profiles with theoretical
predictions.
  If we have underestimated the `true' $M_{500}$ by a
factor of 1.7, we should also have underestimated the `true' $M_{2500}$
by the same factor. This is unlikely: from combined lensing/X-ray
studies, \citet{allen01} conclude that systematic uncertainties are
less than $20$ per cent\footnote{Note that our $M_{2500}$--$T$ relation is
the same, both in slope and normalisation, as that found by
\citet{allen01}.}. Conversely, for the X-ray 
mass profiles to have the correct universal shape, as we have
observed, the predicted difference between the X--ray mass estimates
and the true mass should be roughly constant with radius. This is not
what is expected if the difference is important at $\delta=500$: it is
linked to differences between the temperature profile derived from
projected $\Tsl(r)$ values and the true profile, which depends on the
ICM structure along the line of sight \citep{mazzotta04,rasia05}, and
thus {\it a priori} on radius. It would be interesting to check this  
point
with numerical simulations.

A more likely explanation is that numerical simulations do not
correctly describe the gas thermal structure at large scale, at least
for relaxed clusters considered here.
We note that numerical simulations predict temperature profiles
decreasing  with radius, by nearly a factor 2 at $\delta=500$
\citep[Fig 6]{borgani04}, while the observed profile is flatter
(Fig.~\ref{fig:ktprof}). This would bias low the $\Tsl$ as compared to
the X--ray temperature of real clusters, and thus increase the
normalisation of the $M_{500}$--$\Tsl$ relation \footnote{Since $\Tem$  
is
less affected, the predicted $M_{500}$--$\Tem$ relations would be in
better agreement with observations. }.  It would be interesting to
compare the theoretical $M_{2500}$--$\Tsl$ and $M_{2500}$--$\Tm$
relations, and investigate if there is continued good agreement with
observations.  We expect this to be the case since the predicted
temperature variations are not dramatic -- less than $20$ per cent variations
within $\delta=2500$  (see Fig 6 of \citealt{borgani04} and Fig 9 of
\citealt{kay04a}), so that $\Tsl$ should be close to $\Tm$.

\subsection{The slope of the \MT\  relation}
\label{sec:dissmodslope}

The observed \MT\ relation slope is  consistent  with the self-similar
expectation for the sub-sample of hot clusters  ($T>3.5~\keV$):
$\alpha=1.51\pm0.11$ at $\delta=2500$, where it is best constrained.
The slope is significantly higher when the whole sample ($T>2$~keV) is
considered: $\alpha=1.71\pm0.07$.

A value of $\alpha=1.5$ is expected from the virial theorem if clusters
obey self-similarity. All adiabatic simulations confirm this value
\citep{evrard96,pen98,eke98,bryan98,yoshikawa00,thomas01}, including
when a wide bandpass spectral temperature, as measured with \chandra\
or \xmm , is used to establish the \MT\ relation \citep{mathiesen01}.
Numerical simulations including cooling and feedback do predict a
slightly higher slope. However, the effect is smaller than we observe
($\Delta(\alpha)=0.05-0.1$) and is generally not significant
(Table~\ref{tab:comp}). The only exception is the $M_{500}$--$\Tsl$
relation derived by \citet{rasia05}: $\alpha=1.66\pm0.09$; however the
normalisation is then much too high (see above). It is also worth
noting that the phenomenological analytical model of \citet{voit02}
yields a steeper slope. We obtained $\alpha\sim 1.7$ by fitting their
relation (their Fig 22) in our temperature range. This is in good
agreement with the observed value; however, in this case the
larger slope is mostly due to the variation of the concentration of
the Dark Matter with mass in their model \citep{voit02}, which is
larger than we observe (Paper I).

As a final remark, we want to emphasise that the observed discrepancy
with the standard self-similar value is actually small. The slope
increase, observed when including cool clusters, is significant at most
at the $\sim 85$ per cent confidence level. Furthermore, at  $2~\keV$, the
limiting temperature of our sample, this corresponds to only $-20$ per
cent difference in mass as compared to the extrapolation of the best fitting
\MT\ relation for hotter clusters (see also Fig.~\ref{fig:mt}). There
is scatter in the \MT\ relation, and our sample comprises only 4 cool
clusters. We thus cannot exclude that the steepening is an artefact of
our particular choice of clusters. We also note that  the quality
of the power law fit  decreases when including low mass systems. This
may indicate that  the \MT\ relation is actually convex in the log-log
plane, either across the entire temperature range, or below a 'break'
temperature.  We lack clusters in the intermediate temperature range to
   assess this issue. Clearly, a possible discrepancy between predicted
and observed slopes needs to be confirmed and better specified by
considering a larger cluster sample.

%__________________________________________________________________

\section{Conclusion}
\label{sec:concl}

Using a sample of ten relaxed galaxy clusters observed with \xmm, we
have calibrated the local \MdT\ relation, in the temperature range
$[2-9]~\keV$, at four density contrasts,
$\delta=2500,1000,500,200$. We used the spectroscopic temperature
estimated within $0.5\rv$ ($\delta\sim 1000$), excluding the cooling
core region, and derived the masses at various $\delta$ from NFW
profile fits to precise mass profiles measured up to at least
$\delta=1000$.  We argue that our measured masses are particularly
reliable. The logarithmic slope of the \MdT\ relation is the same at
all $\delta$, reflecting the self-similarity of the mass profiles. The
slope is well constrained and is consistent with the standard
self-similar expectation, $\alpha=1.5$, for the sub-sample of hot
clusters ($T>3.5~\keV$). The relation steepens to $\alpha\sim1.7$ when
the whole sample ($T>2$~keV) is considered. The normalisation of the
\MT\ relation is measured with a precision better than $\pm 5$ per
cent and is $30$ per cent below the value predicted by the adiabatic
numerical simulations of \citet{evrard96}.

Models that take into account radiative cooling and galaxy feedback
are now in good agreement with the observed $M_{2500}$--$T$ relation.
We argue that remaining discrepancies at $\delta=500$ and lower are
more likely to be due deficiencies in models of the ICM thermal
structure, to which the spectroscopic-like temperature seems to be
very sensitive, rather than to an incorrect estimate of the mass from
X-ray data.

More detailed comparisons are needed to understand the origin of the
discrepancies between the predicted and observed \MT\ relations. Our
directly measured $M_{1000}$--$T$ relation now provides the most
direct constraint at large scale for numerical simulations.
Simulations of mass profiles, as would be determined by an X-ray
observer using modern \chandra\ and \xmm\ techniques, are also
needed. This would be particularly interesting for relaxed cluster
sub-samples, and using better representations of observed temperatures
(e.g. as proposed by \citealt{mazzotta04}). Such data could be directly
compared to observed mass profiles.  This would provide information on
i) possible overall systematic errors in X-ray mass estimates, and ii)
further test the reliability of simulations to correctly reproduce the
ICM structure.

On the observational side, study of a much larger, unbiased, sample
is needed to i) determine the exact shape of the local \MT\ relation;  
ii)
study its intrinsic scatter, and iii)  assess the effect of cluster
dynamical state on the \MT\ relation.

%__________________________________________________________________

\begin{acknowledgements}
    We thank A. Evrard for useful comments on the manuscript. We
    thank M. Bershady for providing the BCES software and for
    useful discussions on the statistical analysis. We are grateful to J.
    Ballet and H. Bourdin for further useful discussions on
    statistical analysis. The present work is based on observations
    obtained with \xmm\ an ESA science mission with instruments and
    contributions directly funded by ESA Member States and the USA
    (NASA).  EP acknowledges the financial support of CNES (the French
    space agency). GWP acknowledges funding from a Marie Curie
    Intra-European Fellowship under the FP6 programme (Contract
    No. MEIF-CT-2003-500915).

\end{acknowledgements}


\begin{thebibliography}{63}
\expandafter\ifx\csname natexlab\endcsname\relax\def\natexlab#1{#1}\fi


\bibitem[{{Akritas} \& {Bershady}(1996)}]{akritas96}
{Akritas} M.G. \& {Bershady} M.A 1996, \apj, 470, 706

\bibitem[{{Allen} {et~al.}(2001){Allen}, {Schmidt}, \&
{Fabian}}]{allen01}
{Allen}, S.~W., {Schmidt}, R.~W., \& {Fabian}, A.~C. 2001, \mnras, 328,
L37

\bibitem[{{Bertschinger}(1998)}]{bertschinger98}
{Bertschinger}, E. 1998, \araa, 36, 599

\bibitem[{{Borgani} (2003){Borgani}}]{borgani03}
{Borgani}, S., 2003, in "The Riddle of Cooling Flows in Galaxies and
Clusters of galaxies", Eds. T. Reiprich, J. Kempner, and N. Soker,
http://www.astro.virginia.edu/coolflow/

\bibitem[{{Borgani} {et~al.}(2004){Borgani}, {Murante}, {Springel},
{Diaferio},
    {Dolag}, {Moscardini}, {Tormen}, {Tornatore}, \& {Tozzi}}]{borgani04}
{Borgani}, S., {Murante}, G., {Springel}, V., {et~al.} 2004, \mnras,
348, 1078

\bibitem[{{Bryan} \& {Norman}(1998)}]{bryan98}
{Bryan}, G.~L. \& {Norman}, M.~L. 1998, \apj, 495, 80

\bibitem[{{Castillo-Morales} \& {Schindler}(2003)}]{castillo03}
{Castillo-Morales}, A. \& {Schindler}, S. 2003, \aap, 403, 433

\bibitem[{{Ettori} {et~al.}(2002){Ettori}, {De Grandi}, \&
    {Molendi}}]{ettori02}
{Ettori}, S., {De Grandi}, S., \& {Molendi}, S. 2002, \aap, 391, 841

\bibitem[{{Evrard} {et~al.}(1996){Evrard}, {Metzler}, \&
{Navarro}}]{evrard96}
{Evrard}, A.~E., {Metzler}, C.~A., \& {Navarro}, J.~F. 1996, \apj, 469,
494

\bibitem[{{Evrard} \& {Gioia}(2002){Evrard}, \& {Gioia}}]{evrard02}
{Evrard}, A.E., \& {Gioia}, I. 2002, in Merging Processes in Galaxy
Clusters,
ed.  L. Feretti, I.M. Gioia, G. Giovannini, Astrophysics and Space
Science Library, Vol.  272, 253

\bibitem[{{Finoguenov} {et~al.}(2001){Finoguenov}, {Reiprich}, \& {B{\"
    o}hringer}}]{finoguenov01}
{Finoguenov}, A., {Reiprich}, T.~H., \& {B{\" o}hringer}, H. 2001,
\aap, 368,
    749

\bibitem[{{Eke} {et~al.}(1998){Eke}, {Navarro}, \&{Frenk}}]{eke98}
{Eke}, V.R., {Navarro}, J.F.,  \&{Frenk}, C.~S. 1998, \apj, 503, 569

\bibitem[{{Henry}(2004){Henry}}]{henry04}
Henry, J.P., 2004, ApJ, 609, 603

\bibitem[{{Horner} {et~al.}(1999){Horner}, {Mushotzky}, \&
{Scharf}}]{horner99}
{Horner}, D.~J., {Mushotzky}, R.~F., \& {Scharf}, C.~A. 1999, \apj,
520, 78

\bibitem[{{Kay} {et~al.}(2004a){Kay}, {Thomas}, {Jenkins}, \&
{Pearce}}]{kay04a}
{Kay}, S.~T., {Thomas}, P.~A., {Jenkins}, A., \& {Pearce}, F.~R. 2004a,
    MNRAS, 355, 1091

\bibitem[{{Kay} {et~al.}(2004b){Kay}, {da Silva}, {Aghanim},
{Blanchard}, {Liddle}, {Puget}, {Sadat}, {Thomas}}]{kay04b}
{Kay}, S.~T., {Thomas}, P.~A., {Jenkins}, A., \& {Pearce}, F.~R. 2004b,
astro-ph/0411650.

\bibitem[{{Kotov} \& {Vikhlinin}(2005)}]{kotov05}
{Kotov}, O. \& {Vikhlinin}, A.  2005, \apj, submitted, astro-ph/0504233


\bibitem[{{Mathiesen} \& {Evrard}(2001)}]{mathiesen01}
{Mathiesen}, B.~F. \& {Evrard}, A.~E. 2001, \apj, 546, 100

\bibitem[{{Mazzotta} {et~al.}(2004){Mazzotta}, {Rasia}, {Moscardini}, \&
    {Tormen}}]{mazzotta04}
{Mazzotta}, P., {Rasia}, E., {Moscardini}, L., \& {Tormen}, G. 2004,  
\mnras, 354, 10

\bibitem[{{Muanwong} {et~al.}(2002){Muanwong}, {Thomas}, {Kay}, \&
    {Pearce}}]{muanwong02}
{Muanwong}, O., {Thomas}, P.~A., {Kay}, S.~T., \& {Pearce}, F.~R. 2002,
\mnras,
    336, 527

\bibitem[{{Navarro} {et~al.}(1997){Navarro}, {Frenk}, \&
{White}}]{navarro97}
{Navarro}, J.~F., {Frenk}, C.~S., \& {White}, S.~D.~M. 1997, \apj, 490,
493

\bibitem[{{Neumann} \& {Arnaud}(1999)}]{neumann99}
{Neumann}, D.~M. \& {Arnaud}, M.,1999, \aap, 348, 711

\bibitem[{{Nevalainen} {et~al.}(2000){Nevalainen}, {Markevitch}, \&
    {Forman}}]{nevalainen00}
{Nevalainen}, J., {Markevitch}, M., \& {Forman}, W. 2000, \apj, 532, 694

\bibitem[{{Pen}(1998)}]{pen98}
{Pen}, U.~L.  1998, \apj, 498, 60

\bibitem[{{Pierpaoli} {et al.}(2003){Pierpaoli},{Borgani},{Scott}, \&
{White}}]{pierpaoli03}
{Pierpaoli}, E., {Borgani}, S., {Scott}, D., \& {White}, M.,  2003,
\mnras, 342, 163

\bibitem[{{Pointecouteau} {et~al.}(2004){Pointecouteau}, {Arnaud},
{Kaastra},
    \& {de Plaa}}]{pointecouteau04}
{Pointecouteau}, E., {Arnaud}, M., {Kaastra}, J., \& {de Plaa}, J.
2004, \aap,
    423, 33

\bibitem[{{Pointecouteau} {et al.}(2005){Pointecouteau}, {Arnaud}
\& {Pratt}}]{pointecouteau05}
{Pointecouteau}, E., {Arnaud}, M. \& {Pratt}, G.W., 2005, \aa, 435, 1
(astro-ph/0501635) 

\bibitem[{{Pratt} \& {Arnaud}(2002)}]{pratt02}
{Pratt}, G.~W. \& {Arnaud}, M. 2002, \aap, 394, 375

\bibitem[{{Pratt} \& {Arnaud}(2005)}]{pratt05}
{Pratt}, G.~W. \& {Arnaud}, M. 2005, \aap, 429, 791

\bibitem[{{Press} {\etal}(1992)}]{numrec}
{Press}, W.H., {Teukolsky}, S.A., {Vetterling}, S.A. \& {Flannery}  
B.P.,  1992, Numerical Recipes in Fortran 77, Second Edition, p.660

\bibitem[{{Rasia} {et~al.}(2004){Rasia},
    {Tormen},  \& {Moscardini}}]{rasia04}
{Rasia}, E., {Tormen}, G., \& {Moscardini}, L., 2004, \mnras, 351, 237

\bibitem[{{Rasia} {et~al.}(2005){Rasia}, {Mazzotta}, {Borgani}, L.,
{Dolag},
    {Tormen}, {Diaferio}, \& {Murante}}]{rasia05}
{Rasia}, E., {Mazzotta}, P., {Borgani}, S., {et~al.} 2005, \apj, 618, L1

\bibitem[{{Rowley} {et~al.}(2004){Rowley}, {Thomas}, \&
{Kay}}]{rowley04}
{Rowley}, D.~R., {Thomas}, P.~A., \& {Kay}, S.~T. 2004, \mnras, 352, 508

\bibitem[{{Sanderson} {et~al.}(2003){Sanderson}, {Ponman}, {Finoguenov},
    {Lloyd-Davies}, \& {Markevitch}}]{sanderson03}
{Sanderson}, A.~J.~R., {Ponman}, T.~J., {Finoguenov}, A.,
{Lloyd-Davies},
    E.~J., \& {Markevitch}, M. 2003, \mnras, 340, 989

\bibitem[{{Thomas} {et~al.}(2001){Thomas}, {Muanwong},{Kay}, {Pearce},
{Couchman}, {Edge}, {Jenkins} \&
    {Onuora}}]{thomas01}
{Thomas}, P.~A., {Muanwong}, O.,  {Kay}, S.~T., {Pearce},
F.~R.,{Couchman}, H.M.P., {Edge}, A.C., {Jenkins} A., \&
    {Onuora} L., 2001, \mnras,
   324, 450

\bibitem[{{Thomas} {et~al.}(2002){Thomas}, {Muanwong}, {Kay} &
{Liddle}}]{thomas02}
{Thomas}, P.~A., {Muanwong}, O.,  {Kay}, S.~T., \& {Liddle}, A.R., ,
2002, \mnras,  330, L48

\bibitem[{{Viana} {et~al.}(2003){Viana},{Kay},{Liddle}, {Muanwong} \&
{Thomas}}]{viana03}
{Viana}, P., {Kay}, S.T., {Liddle}, A.R., {Muanwong}, O.,  \& {Thomas},
P.A., 2003, \mnras, 346, 319

\bibitem[{{Voit} {et~al.}(2002){Voit}, {Bryan}, {Balogh} \&
{Bower}}]{voit02}
           {Voit}, G.M., {Bryan}, G.L., {Balogh}, M.L., \&  {Bower}, R.G.
2002, ApJ, 576, 601

\bibitem[{{Voit} \& {Ponman}(2003)}]{voit03}
         {Voit}, G.M., {Ponman}, T.J., 2003, ApJ, 594, L75

\bibitem[{{Xu} {et~al.}(2001){Xu}, {Jin}, \& {Wu}}]{xu01}
{Xu}, H., {Jin}, G., \& {Wu}, X.~P. 2001, \apj, 553, 78

\bibitem[{{Yoshikawa} {et~al.}(2000){Yoshikawa}, {Jing}, \&
{Suto}}]{yoshikawa00}
{Yoshikawa}, K., {Jing}, Y.P.,  \& {Suto}, Y., 2000, \apj, 535, 593





\end{thebibliography}
\end{document}